\documentclass{IOS-Book-Article}

\usepackage{courier}
\usepackage{graphicx}
\usepackage{listings}
\usepackage{mathptmx}

\usepackage{program}
\usepackage{seqsplit}
\usepackage{afterpage}
\usepackage{algorithm}
\usepackage{url}
\usepackage{longtable}
\usepackage{textcomp}
\usepackage{enumitem}

\def\hb{\hbox to 10.7 cm{}}

\begin{document}

\pagestyle{headings}
\def\thepage{}

\begin{frontmatter}              

\title{OK Google, What Is Your Ontology? \\Or: Exploring Freebase Classification to Understand Google's Knowledge Graph}

\markboth{}{\hb}

\author[A]{\fnms{Niel Chah}%
\thanks{University of Toronto, Faculty of Information
niel.chah@mail.utoronto.ca}}

\runningauthor{N. Chah}
\address[A]{University of Toronto, Faculty of Information}

\begin{abstract}
This paper reconstructs the Freebase data dumps to understand the underlying ontology behind Google's semantic search feature. The Freebase knowledge base was a major Semantic Web and linked data technology that was acquired by Google in 2010 to support the Google Knowledge Graph, the backend for Google search results that include structured answers to queries instead of a series of links to external resources. After its shutdown in 2016, Freebase is contained in a data dump of 1.9 billion Resource Description Format (RDF) triples. A recomposition of the Freebase ontology will be analyzed in relation to concepts and insights from the literature on classification by Bowker and Star. This paper will explore how the Freebase ontology is shaped by many of the forces that also shape classification systems through a deep dive into the ontology and a small correlational study. These findings will provide a glimpse into the proprietary blackbox Knowledge Graph and what is meant by Google's mission to ``organize the world's information and make it universally accessible and useful''.
\end{abstract}

\begin{elskeyword}
ontology \sep Google \sep Freebase \sep Knowledge Graph \sep classification
\end{elskeyword}
\end{frontmatter}
\markboth{\hb}{\hb}

\section{Introduction}

\begin{figure}[h]
	\centering
	\caption{A Knowledge Panel in Google Search}
	\includegraphics[width=.75\columnwidth]{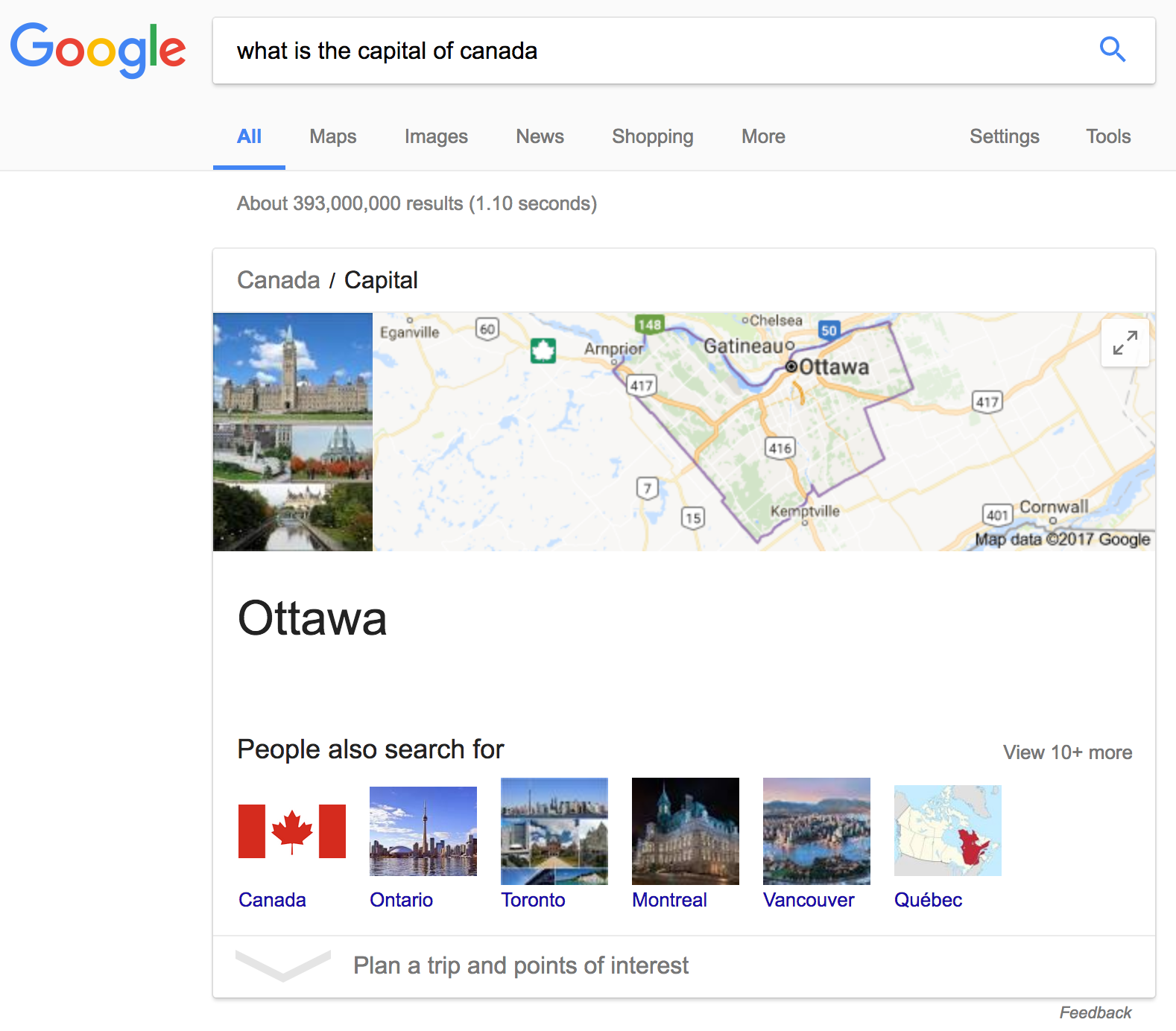}
	\label{fig1}
\end{figure}

This paper's title is a riff on a key phrase ``OK, Google'' that is uttered by users to trigger Google's virtual assistant services. Through products such as the Google Home appliance and a mobile application, the Google Assistant taps into the structured data in the proprietary Google Knowledge Graph (KG) to answer queries like ``What is the capital of Canada?''. The KG can be most prominently seen in the structured answers to queries in the web search engine (see Figure \ref{fig1}). The underlying structures in the KG that allow Google to offer this significantly different service from its original offerings in providing links to external resources will be explored.

This paper will consider the underlying ontology that structures the Knowledge Graph and by extension the kinds of answers that Google Search or Assistant can provide. By reconstructing the Freebase knowledge base that was acquired by Google to power its Knowledge Graph, this paper will provide a glimpse into the proprietary and blackbox system that is used by millions of users for information retrieval and question answering. The structures found in the Freebase/Knowledge Graph ontology will be analyzed in light of the findings on classification systems in a key text by Bowker and Star (2000) \cite{bowker2000sorting}.

\section{Background: Freebase and Google Knowledge Graph}

\begin{figure}[h]
	\centering
	\caption{A screenshot of freebase.com on May 2, 2016 before it was shut down.}
	\includegraphics[width=.75\columnwidth]{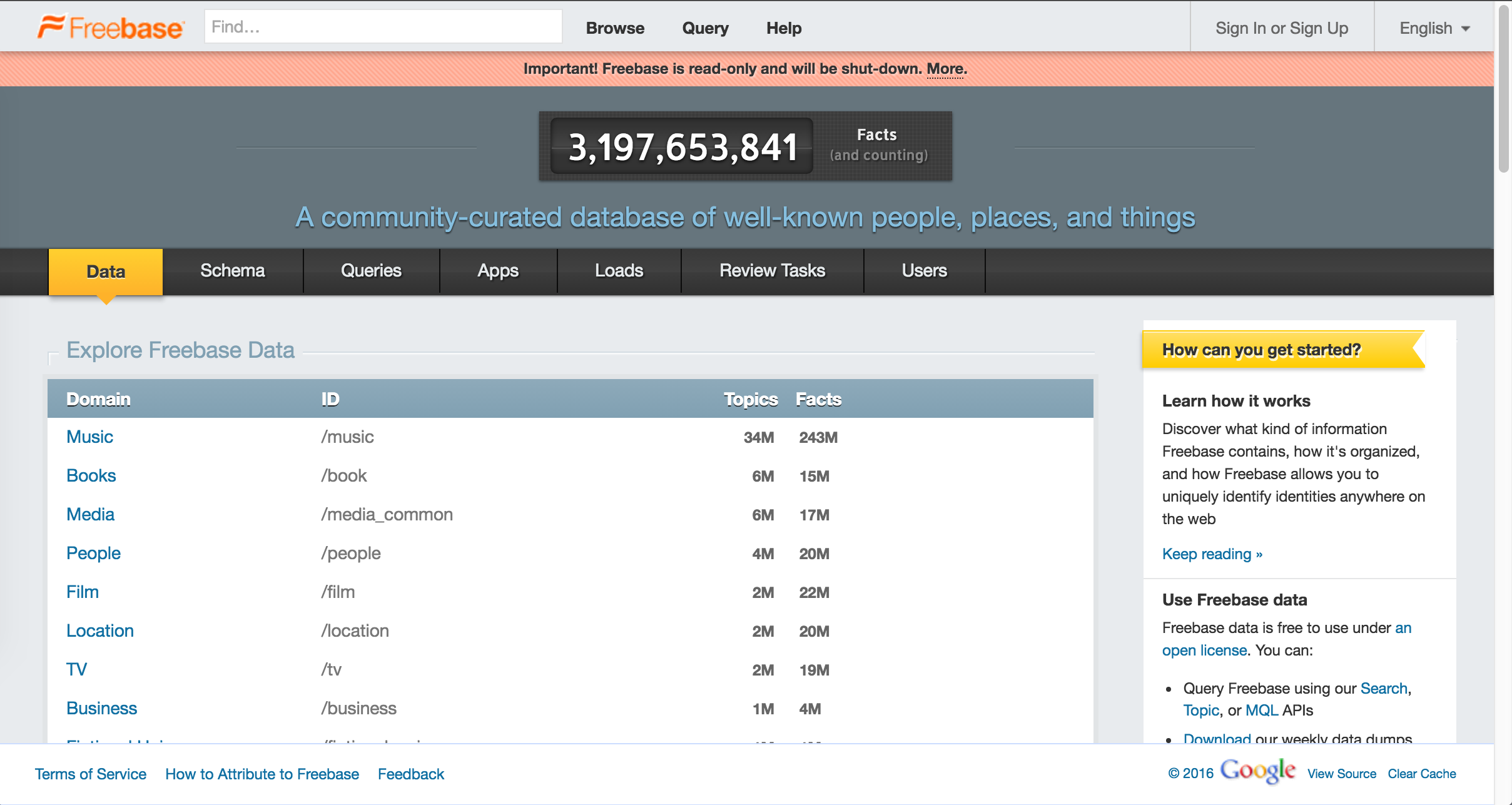}
	\label{fig2}
\end{figure}

In 2007, Freebase, available at freebase.com (see Figure \ref{fig2}), was launched by Metaweb as an open and collaborative knowledge base \cite{Bollacker2008}. On the website, users could register a free account and edit the data of various entries, creating linkages between entities in the data. This was similar to the kind of editing and data entry that was possible on Wikipedia. A significant difference was that while Wikipedia predominantly consists of free-form text suitable for an article, Freebase encoded links and relationships between entities in the knowledge base.

In 2010, Metaweb was acquired by Google for an undisclosed sum \cite{Menzel2010}. Freebase data was then used to power their internal Knowledge Graph to support Google Search features such as the Knowledge Panels (see Figure \ref{fig1}) \cite{PellissierTanon2016}. On December 16, 2014, it was announced on the Freebase Google Plus page that Freebase would be gradually shut down over the next six months, and its data would be migrated to Wikidata \cite{Gplus2017}. In reality, the Freebase service was closed from any further edits by its users and finally shut down on May 2, 2016 \cite{Douglas2016}. From this date onward, all freebase.com URLs began to redirect to the Google Developers page for the project where a data dump of Freebase data is available.

\section{Definition of Terms}

It is important to define the concepts and vocabulary that will be used throughout this paper. An attempt has been made to start with the most indivisible parts, and build a vocabulary from them below. In Table \ref{table1}, an example of a thing in the real world, a conceptual counterpart, and its implementation in Freebase have been mapped together. After each term has been defined below, the Freebase term will be used throughout the paper for consistency.

\begin{center}
\begin{table}[h]
	\caption{Equivalent Terms in Defining Parts of an Ontology}
	\label{table1}
\begin{tabular}{ |p{2cm}|p{2cm}|p{2cm}|p{4cm}| }  
\hline
\textbf{Example from the World} & \textbf{Concept} & \textbf{Implementation in Freebase} & \textbf{Comment} \\
\hline
Barack Obama & Entity or Instance & Freebase Object & The specific U.S. president. \\
\hline
Person & Class & Freebase Type & Obama belongs to the class of ``people''.  \\
\hline
Place of Birth & Relation or Property or Slot & Freebase Property & The ``place of birth'' relation links Obama and Hawaii. \\
\hline
Place of Birth accepts object of Location type & Facet & Freebase Property Detail & The ``place of birth'' relation links Obama and Hawaii. Hawaii is an object with type Location.\\
\hline
Barack Obama was born in Hawaii. & Fact & RDF Triple & ``Fact'' is used in the same sense as ``statement'' or ``assertion''.\\
\hline
\end{tabular}
\end{table}
\end{center}

\textbf{\textit{Object}} - A Freebase object is a globally unique identifier that is a representation in Freebase of something in the world. This broad definition allows arbitrary things to be represented in the data. A tangible thing, such as the person Barack Obama, or an abstract thing, such as the number ``5'', can be represented by a Freebase object. An implementation detail is that a Freebase object is written as a unique machineId (a \textit{mid}) that is in the format ``/m/'' + \{alphanumeric\}. For instance, ``/m/abc123'' is the Freebase object for the person Barack Obama. Further interesting things can then reference the ``/m/abc123'' identifier to express interesting relations or properties of Barack Obama.

\textbf{\textit{Type}} - A Freebase type is used to express the concept of a class. According to Freebase idiomatic usage, it is said that a Freebase object ``has \textit{X, Y, Z} types''. This is equivalent to saying that such an object ``is a member of the \textit{X, Y, Z} classes''. In the case of U.S. President Barack Obama, the Freebase object for Obama would have the types ``/people/person'' and ``/government/us\_president'' (the exact notation of these types will be explained later) to indicate that the object is a member of the Persons and U.S. Presidents classes.

\textbf{\textit{Property}} - A Freebase property is a relation that describes how an object can be linked to other values or objects. A property can link an object to a value in the case of a person's name value ``Barack Obama'' or a date of birth value ``1960''. A property that links an object to another object is the ``place of birth'' as it links an object of People type with an object of Location type. An implementation detail is that a Freebase property is situated in the context of a type (i.e., the People type specifies a Place of Birth property, while other types specify different properties). This will be briefly explained in a short comment following these definitions.

\textbf{\textit{Property Detail}} - A property detail refers to the constraints on what objects or values can be linked through a property. A property can link an object and a value in the case of a person's name value ``Barack Obama'' or a date of birth value ``1960'', or it can link an object and another object in the case of a person's place of birth ``(the object for) Hawaii'' The specifications that ``the name property links to a text value'' or that ``the place of birth accepts a Location typed object'' are property details in Freebase.

\textbf{\textit{RDF triple}} - The Resource Description Format (RDF) is a specification for data representation in a ``triples'' (or tuple of N=3) format \cite{RDF2014}. An RDF triple is a single line of data made up of 3 ordered parts: the \textit{subject}, \textit{predicate}, and \textit{object}. Each part of the triple plays a role in the data expressed. From left to right, the (1) \textit{subject}, through the (2) \textit{predicate} part, expressed a relationship to the (3) object. This format intuitively mirrors how a single triple expresses a fact through a sentence-like format: Johnny Appleseed (subject) likes (predicate) apples (object).

\textbf{\textit{Ontology}} - For this paper, the Freebase \textit{ontology} is the formal structure and description of the \textit{types}, \textit{properties} and \textit{property details} that specify how \textit{objects} can be related to one another. This definition adapts the definition by Noy and McGuinness that ``an ontology is a formal explicit description of concepts in a domain of discourse (classes (sometimes called concepts)), properties of each concept describing various features and attributes of the concept (slots (sometimes called roles or properties)), and restrictions on slots (facets (sometimes called role restrictions))...An ontology together with a set of individual instances of classes constitutes a knowledge base'' \cite{noy2001ontology}. An implementation detail is that this ontology is also expressed as RDF triples (e.g. by saying ``\textit{Place of Birth - is a - property}''). 

Noy and McGuiness state there is a distinction between ``the ontology'' and ``the set of individual instances of classes'', but also advise that in practice ``there is a fine line where the ontology ends and the knowledge base begins''\cite{noy2001ontology}. For example, as Obama is an instance of the Person and US President types (classes), the RDF triples that express this would be considered as part of the knowledge base and not the ontology. It is not just instances of classes that can be expressed. A Freebase object for the Honda Civic would represent not a single instance of the automobile, but an abstracted notion of Honda Civics in general. Depending on how this is expressed, the ontology could include the notion of Honda Civics as a Freebase type (a class) where Obama's Honda could be an instance of this class.

\textbf{\textit{Architecture}} - In this paper, the architecture refers to the kind of general patterns and relationships that can be found in the ontology. Does the ontology allow for inheritance between classes (or types in Freebase parlance)? Are there default values associated with properties? How are ``zero'' or null values handled? These kinds of questions that are not necessarily concerned about what is specifically expressed in the ontology (whether Planes, Trains, or Automobiles) but more so about \textit{the ways in which} the ontology expresses it  should be addressed by examining the architecture.

\hfill

The Freeabse ontology of types and properties also has an interesting human-readable aspect that borrows from the Unix-like use of forward slashes ``/'' in directories. In the Freebase ontology, there are a number of top-level \textit{domains} on specific subject matters. For example, there is a domain for People and another domain for Films, expressed as the ``/people'' and ``/film'' domains respectively. Within each domain, there are ``type(s)'' that represent the classes of entities in a domain. Thus, there is a Person type, expressed by a human-readable ID as ``/people/person'', and a Deceased Person type, expressed as ``/people/deceased\_person'' type. In the Film domain, notable types include the Film Director ``/film/film\_director'' and Film ``/film/film''. Within each type in turn, there are properties to capture the granular facts, such as the Date of Birth for people ``/people/person/date\_of\_birth''.

\section{Methodology}

The data collection and pre-processing process will be briefly explained. The Freebase data dump of 1.9 billion triples is available for download in a N-Triples RDF format \cite{FreebaseDumps}.\footnote{freebase.com} The data dumps were parsed using shell scripts and command line tools run on a MacBook Pro (Early 2015 model, 2.7 GHz Intel Core i5, 8 GB RAM) and an external hard drive (Seagate 1 TB) \cite{Chah2017}. Open source Unix-like tools such as awk, cut, grep, less, more, parallel, sed, sort, wc, zless, zmore, and zgrep were used to slice out portions of the data based on the Freebase ontology. 

A slice is defined as a subset of the RDF triples data, where the triple's \textit{predicate} (the middle term or ``verb'' in a simple sentence) is part of a unique domain, type, or property. For example, a slice for all of the RDF triples that have a predicate in the form ``/people/*'' (where the asterisk represents a wildcard) are put into a distinct slice. This ``people'' slice then expresses all data on people-typed properties. This is an intuitive slicing method since the predicate term is the edge or link between the subject and object nodes, if triples are conceptualized as a graph. With the resulting cleaned data sets, the ontology of various domains were examined in detail. A sample of the slices is shown in Table \ref{table3}. 

\begin{center}
\begin{table}[h]
	\caption{A Selection of Freebase Slices by Domain, ordered alphabetically}
	\label{table3}
\begin{tabular}{ |l|l|r|r| } 
\hline
Slice Name & Predicate & Number of Triples & \% of All Triples \\
\hline
american\_football & /american\_football/* & 278,179 & 0.009 \\
\hline
amusement\_parks & /amusement\_parks/* & 22,880 & 0.001 \\
\hline
architecture & /architecture/* & 253,718 & 0.008 \\
\hline
astronomy & /astronomy/* & 556,381 & 0.018 \\
\hline
automotive & /automotive/* & 46,543 & 0.001 \\
\hline
\end{tabular}
\end{table}
\end{center}

\section{Freebase Ontology and Classification}

Users of the Google search engine and more recently the expanding suite of voice-assisted virtual assistant services in the form of Google Home will probably find great utility in receiving the structured answers to certain queries found in the Knowledge Panels (KPs) or Knowledge Cards. In contrast to the information retrieval services that were the focus of the search engine until only recently, KPs presumably service the user's needs more efficiently, by reducing the user's hops to other external links, and pose a competitive advantage to Google, by keeping users within Google's virtual properties longer. However, as ``convenient'' as it may be to find out at a moment's notice, which team won the latest MLB game or when a celebrity was born, it is important to consider the backend infrastructure that delivers these answers. A certain order or classification is imposed on some data somewhere to implement this service. As Bowker and Star note, ``Information infrastructure is a tricky thing to analyze...the easier they are to use, the harder they are to see.'' \cite{bowker2000sorting}. 
What does the system make sense of? What is left out? What is privileged and by extension what is ignored by Google?

In \textit{Sorting Things Out}, Bowker and Star note how classification schemes are inconsistent, contradictory, not natural, and not universal \cite{bowker2000sorting}. Although the Freebase ontology may not immediately seem like a classification system, the structure of types (classes) and properties is a system based on classifying diverse kinds of things. The Freebase ontology is more complex than the list of animals created by the Chinese emperor as recounted by Borges, but there are still ways in which Freebase is also not ``natural, eloquent, and homogeneous'' \cite{borges1981analytical,bowker2000sorting}. 
As a system for ordering and classifying representations of things in the world, the Freebase ontology will be discussed in light of Bowker and Star's findings on classification. 

Bowker and Star make a distinction between two classical kinds of classification systems: Aristotelian and prototype classification \cite{bowker2000sorting}. Although the distinction between the two is not absolute, it is useful to keep in mind the broadly different approaches in the two systems. An Aristotelian classification operates ``according to a set of binary characteristics that the object being classified either presents or does not present'' while a prototype classification considers ``a broad picture in our minds of what a chair is; and we extend this picture by metaphor and analogy'' \cite{bowker2000sorting}. 
The two main reasons cited by Bowker and Star as leading to the convergence and mixed usages of Aristotelian and prototypical approaches are applicable to this paper. An inconsistency in classification systems arises because (1) ``each classification system is tied to a particular set of coding practices'' and (2) ``classification systems in general...reflect the conflicting, contradictory motives of the sociotechnical situations that gave rise to them'' \cite{bowker2000sorting}. 
The Aristotelian and prototypical characteristics of the Freebase ontology will be pointed out later in the discussion.

In the next sub-sections, this paper will explore a selection of the ontology and architecture of Freebase that is most applicable to Bowker and Star's critiques of classification systems. 
A review of the Freebase type system for expressing entities as members of classes will show how Aristotelian classification processes can be seen. The notion of \textit{incompatible} types will be explained, followed by a reflection on the lack of \textit{inheritance} in the system.
Then, the interesting implementation of Freebase's ``Has Value'' and ``Has No Value'' notation will be explored to see how the system is sufficient for Google's KG usage.
An exploration of how Freebase handles data duplication and data conflation is used to point out how the system is made to serve Google's particular needs. Lastly, a small correlational study is conducted to explore a research question inspired by Bowker and Star.

\subsection{Freebase's Type System}

An interesting observation on the Aristotelian nature of the Freebase ontology can be found from parsing how types (classes) are referred to idiomatically in Freebase.\footnote{Another interesting Freebase idiom that is generally confusing for outsiders is how a Freebase object is referenced in practice. The Freebase object that represents a real world entity is said to be a \textit{topic} (i.e., ``/m/abc123 is the object for the Barack Obama \textit{topic}.''). This jargon was avoided for this paper.}
A Freebase object that represents a real world entity is said to ``have certain \textit{X, Y, Z} types'' to indicate that it is a member of \textit{X, Y, Z} classes. Under this system, types are something that an entity ``has'', much like an entity ``has'' a certain characteristic like size or colour. The film \textit{The Terminator} (1984) had a director, a runtime, and a ``\textit{Film}-ness'' (but not ``\textit{TV Program}-ness'') according to this conceptualization. In this way, types are added like binary descriptors to the entity being represented, indicating that this is a Person, Film, or Event. 

\begin{figure}[h]
	\centering
	\caption{An unfortunate KP that displays the country of Belarus as a book (Aug. 2016).}
	\includegraphics[width=\columnwidth]{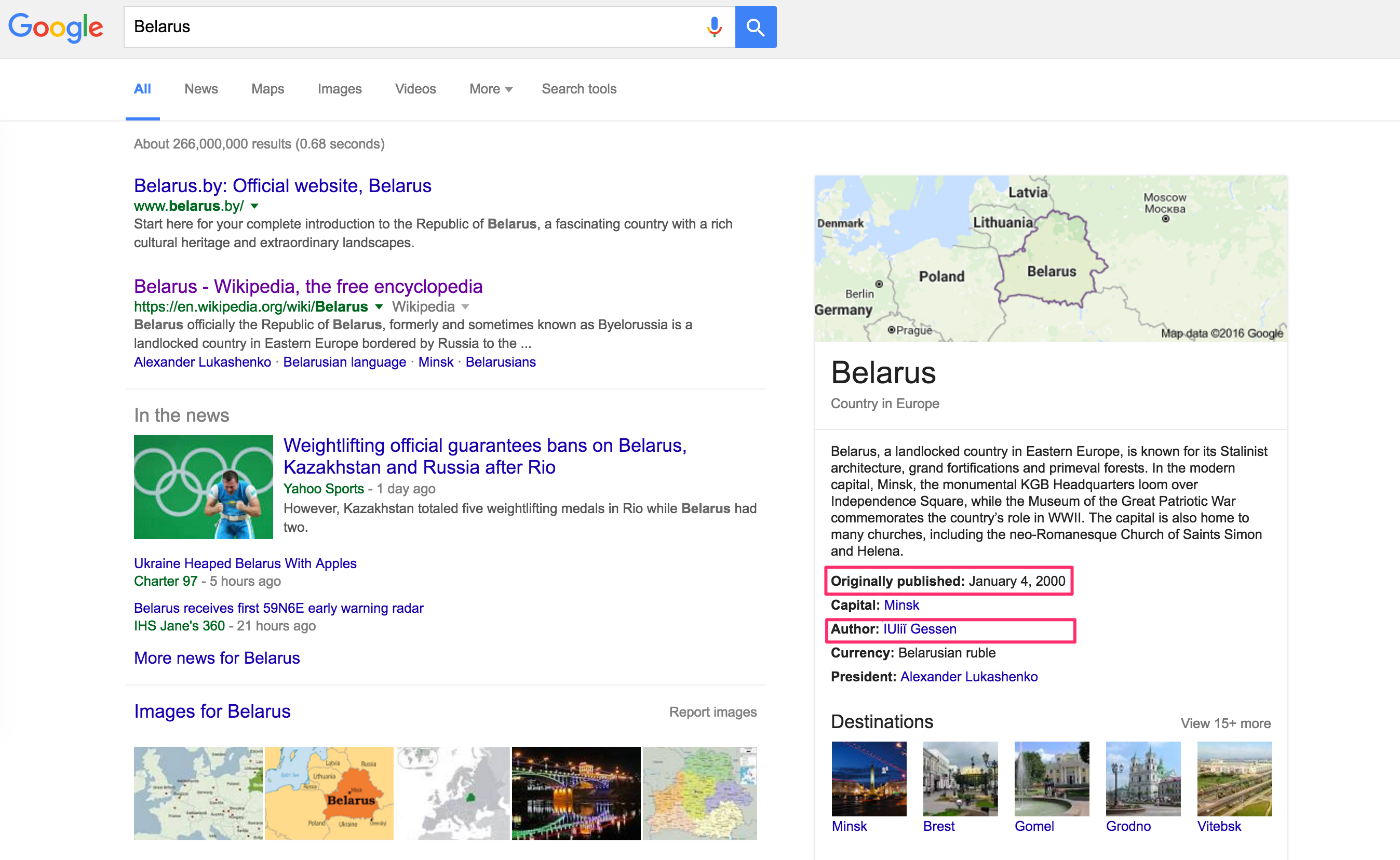}
	\label{fig3}
\end{figure}

The notion of \textit{incompatibility} arises in the Freebase system to express how an object can have certain types which necessarily exclude it from having other types. This is implemented at the RDF triples level (e.g. by saying ``\textit{X - is incompatible with - Y}'') and applies to the whole system. For instance, incompatibility is enforced so that the object representing the film \textit{The Terminator} (1984) and the object representing the \textit{Terminator} film series remain distinct. Without this rule, inaccurate relationships may be expressed and cause embarrassing results in public KPs. For instance, in saying ``\textit{Terminator 2} (1991) - is the Sequel to - \textit{Terminator}'' it is important that this \textit{Terminator} is the Film. The implementation of \textit{incompatibility} in the Freebase architecture serves Google well to prevent these and potentially more impactful mistakes from surfacing in the Knowledge Panels. When there are failures (see Figure \ref{fig3}), this can likely be attributed to the lack of an incompatibility rule.

Unlike many widely known classification systems or even Object Oriented Programming (OOP) language infrastructures, Freebase does not implement \textit{inheritance} in its types. Although there is a Person type and a Film Actor type in Freebase, the latter does not inherit from the former. One possible explanation for the lack of inheritance in the architecture is that the aforementioned incompatibilities serve a sufficiently robust role in ensuring the data does not express an embarrassing, harmful, or incorrect statement that has the potential to be served in a Google KP.

The lack of inheritance may also be a feature that allows entities to be represented with greater flexibility. For instance, if the cast in a film is represented through the Film Actor type, the lack of inheritance makes it relatively easy to express that a famous non-human animal, like a golden retriever, was a Film Actor starring in a film. In the interests of Google Search and KPs, displaying Air Buddy (the dog) as part of the cast of \textit{Air Bud} (1997) may be more important than whether the ontology expresses the structure that Film Actor inherits from Person, or Animal, or Thing. 

A general review of Freebase's type system has been conducted to show how aspects of the architecture (incompatibility, inheritance) and the ontology (how a type is determined and ``had'') are largely shaped by and aligned with Google's interests in presenting data that would be serviceable for users of Knowledge Panels. This observation follows Bowker and Star's characterization of classification systems as being formed for specific uses under specific sociotechnical contexts.

\subsection{Has Value or Has No Value?}

It has been mentioned in passing throughout the paper that the ontology and knowledge base are implemented through RDF triples. Triples can express a fact about an object by linking said object to a value (e.g. by saying ``Canada - was founded in - 1867''). However, how can a triple express an estimated, uncertain, or null value? There must be a way to express ``unknown'' (there is some value, but it is either not known or uncertain) and ``null'' (there is definitely not a value; it is undefined) values in Freebase and the Knowledge Graph. These are called ``Has Value'' (HV) and ``Has No Value'' (HNV) value notations respectively.

In a sense, the HV and HUV notations are ``hacks'' or workarounds to express uncertainty and undefined values while preserving the overall implementation of linked data via RDF triples. First, it is possible to consider a simple implementation of HV and HNV by letting the third term in a triple express the uncertainty as a string value (e.g. by saying ``\textit{Plato - Date of Birth - Unknown Value}''). A slightly more sophisticated implementation may link to an object representing unknown-ness, say /m/unkn0wn. The actual implementation is more roundabout. HV and HUV are expressed by linking the property to the object (e.g. by saying ``\textit{Date of Birth - Has Value - Plato}''). From the data dumps, the descriptions for these HV and HUV properties also state ``Note: this property takes as MID as value. The object has a value for this property, but the actual value isn't known.''\footnote{The precise Freebase notation for these properties are ``/freebase/valuenotation/has\_no\_value'' and ``/freebase/valuenotation/has\_value''. From the data dumps, the triple that expresses the description is \textit{/freebase/valuenotation/has\_value - /common/topic/description - "Note: this property takes as MID as value. The object has a value for this property, but the actual value isn't known. (E.g., we know that this marriage has ended, but we don't know the date it ended.) This property is asserted as a 'bare' property (without /freebase/valuenotation being asserted as a type)."}. The same description applies for /freebase/valuenotation/has\_no\_value.}

\begin{figure}[h]
	\centering
	\caption{A KP that displays ``known'' facts about Plato (Dec. 2017).}
	\includegraphics[width=\columnwidth]{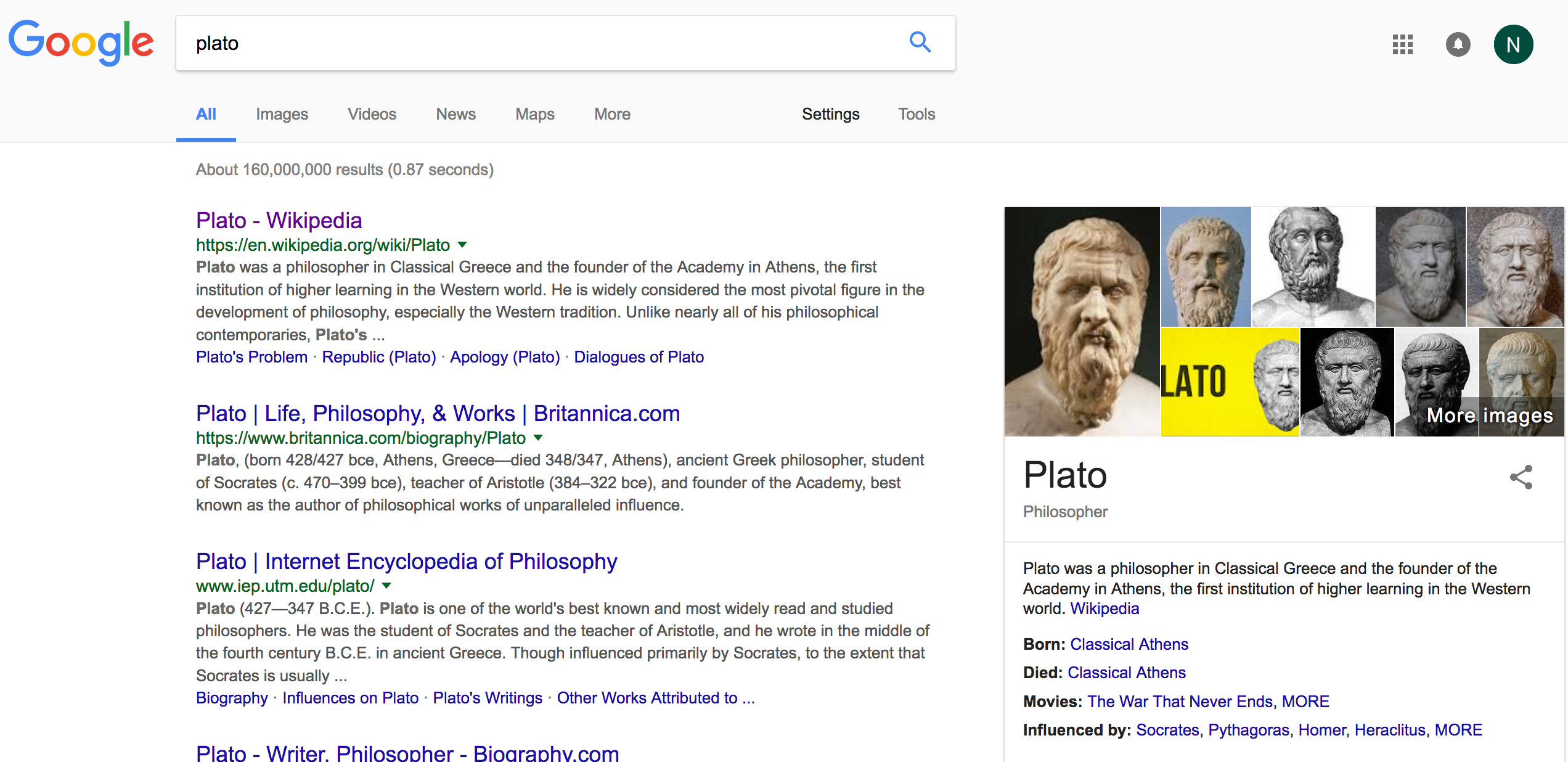}
	\label{fig4}
\end{figure}

The interesting implementation of expressing unknowns and nulls in this way may suggest that Freebase/KG was not initially built to support this level of uncertainty; it was an afterthought once the basic system of triples as a Knowledge Graph was found to work. In hindsight, it may seem like an unwieldy workaround, but this kind of architecture may be exactly what Google finds useful for its purposes. Google KPs are meant to display facts about entities (the ``knowledge'' in Knowledge Panels) and rarely are unknowns or nulls expressed (see Figure \ref{fig4}). The Freebase/KG system expresses known facts well enough for Google to use it in its production environments in Search, Home, and Assistant. The notion that Google's data encodes certain uncertainties is not exposed to the end user although it is certainly implemented in this unique way.

It is also worth considering the level at which uncertainty is expressed in the architecture. There is the notion of Has Values and Has No Value. Values outside of these two possibilities are not supported. A person's Date of Birth could be known as being either 1900 or 1901 (due to a clerical error or a lost record). A book's author could be a well known writer or a group of unknown writers. A time period could have a range of arbitrarily different values ``1901-10'' and ``1930-35''. In all of these cases, Google's system cannot express these details and only presents two possible options.

\subsection{Dealing with Doppelgangers and Chimeras}

This section's colourful title concerns how Freebase deals with ``merging'' duplicate objects (doppelgangers) and ``splitting'' conflated objects (chimeras). A duplicate object refers to another object, ``/m/xyz123'', that represents the same entity as an already existing object, ``/m/abc123''. Duplicates can be a pervasive problem as many data sources, editors, or human contributors can create another unique object to represent an entity with an existing representation in the data. These many objects should be ``merged'' into one. A conflated object can occur when a single object is linked to facts that represent more than 1 entity. This single object should be ``split'' into many. Without these recommended actions, these situations have the potential to lead to embarrassing and damaging representations in Google's KPs.

\begin{figure}[h]
	\centering
	\caption{A KP colourfully marked up by NYT reporter Rachel Abrams (Dec. 2017) \cite{abrams2017nytimes}.}
	\includegraphics[width=.5\columnwidth]{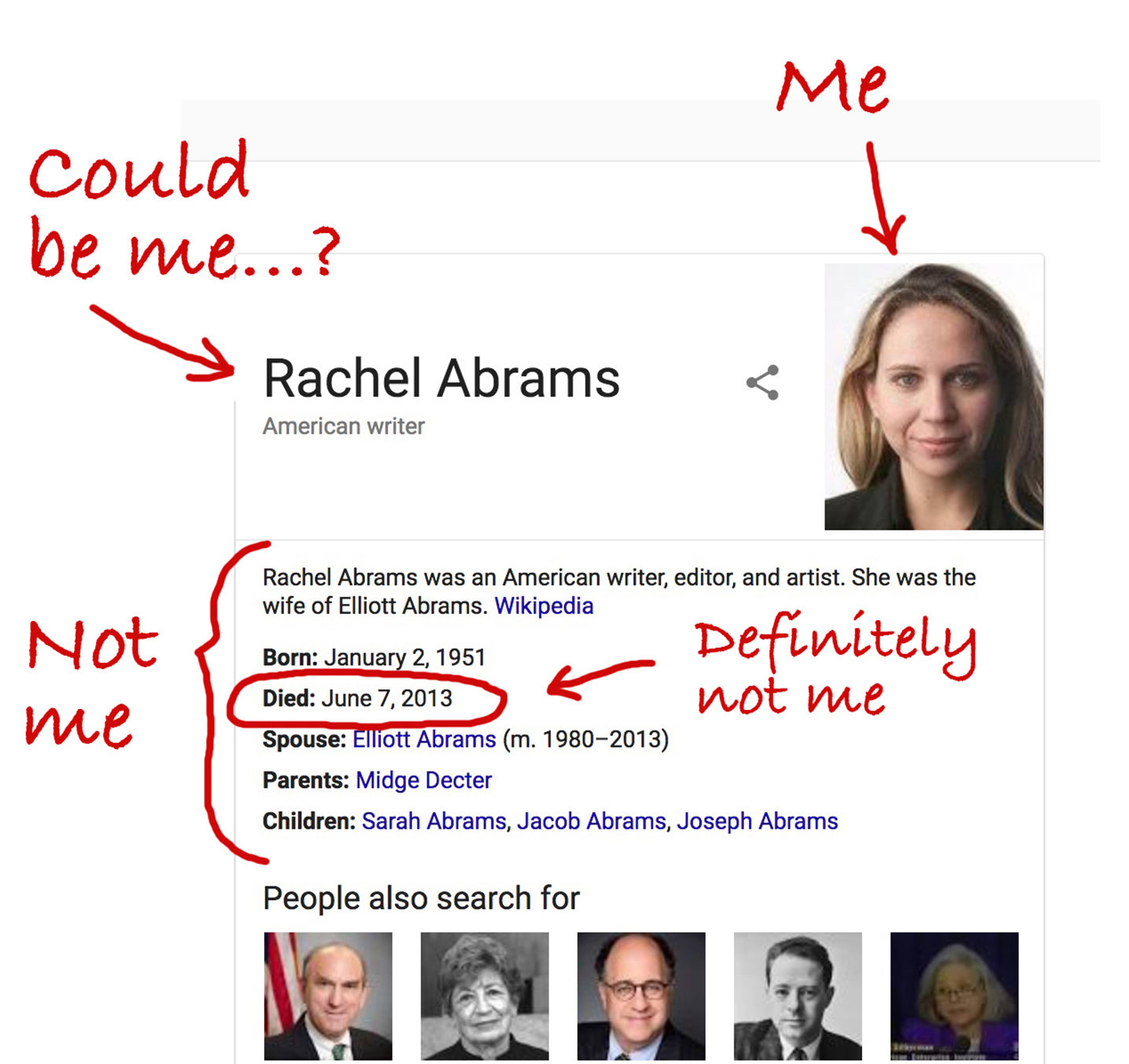}
	\label{fig5}
\end{figure}

The \textit{New York Times} business reporter Rachel Abrams discovered the problems associated with merging and splitting firsthand through a KP that displayed her Date of Death was five years ago (see Figure \ref{fig5}). \cite{abrams2017nytimes}. What occurred here was that the Google KG had merged data on the living reporter Rachel Abrams with another deceased Rachel Abrams. Abrams was in a position to publish an article about her journey in addressing the issue as an end user (and also the very subject of the KP), but the underlying merge/split operations were hidden inside the blackbox of Google's KG.

Just as the ``/freebase/*'' part of the ontology was used to implement HV and HNV value notations, the ``/dataworld/*'' ontology is used to implement merges.\footnote{The description of the ``/dataworld/*'' domain from the data dumps reads as ``Dataworld is a domain for schema that deals with operational or infrastructural information in relation to the knowledge graph. It contains schema that captured information in relation to bulk loading or transformations of graph data.''.}
Specifically, the property ``/dataworld/gardening\_hint/replaced\_by'' is used to implement merges between various objects (e.g. by saying ``\textit{/m/xyz123 - Replaced By - /m/abc123}''). The description of this property from the data dumps states ``This property is used to point to a node that subsumes the identity of this node, usually as a result of a merge operation.'' (where node refers to the term \textit{object} used in this paper). This system of ``Replaced By'' arrows to point to the canonical object is arguably quicker than the alternative of editing each triple where the duplicate object is involved. Like the system of HV and HNV notations, this merging and splitting system works well enough for quickly triaging any Abrams-like mistakes.

In the backend of Freebase/KG, the mechanisms that support merging and splitting objects are important and necessary considering how Google uses the data in such publicly visible ways. A variety of diverse sources (some programmatic and others manual) contribute data to the KG. In the event of a duplicate object being created or an object becoming conflated with another entity, the effects can be readily seen on Google's KPs. It is in Google's interests to adopt an ontology and architecture that can facilitate such fixes.

\section{A Small Correlational Study}

A small correlational study was conducted to explore the following research question. \textit{Is there a relationship between the complexity of a domain's ontology (the types, properties, etc. for the People, Films, etc. domain) and the amount of triples expressing facts (the ``knowledge base'') associated with that ontology?} Phrased another way, are the ontologies for domains that Google actively uses for its KPs, like music, films, and people domains, more ``mature'' than the ontologies for domains that exist in Freebase but are not exposed in KPs? This question is based on a hunch that actively curated and administered  domains for Google's uses have more labour and detail put into them than other less active domains. This aligns with Bowker and Star's observation that classification systems are built for specific uses according to sociotechnical circumstances. For this study, ``complexity'' and ``maturity'' are operationalized by considering the amount of property details (how many descriptions, constraints, etc.) associated with a domain. 

As explained earlier in the Methodology section, the data dumps were transformed into a series of \textit{slices}. Each slice contained a subset of the RDF triples where the predicate (middle) term corresponded to the domain. The People slice contains all triples where the predicate is of the form ``/people/*''. The resulting domains for each slices were examined and organized into three large categories (the full table is available in Appendix A). The domains that (1) implement Freebase features at a technical level and (2) duplicate triples through the OWL ontology were ignored.

This left 89 slices on domains concerning specific subject matters, such as music, films, and TV. As mentioned earlier in the paper, the ontology itself is also expressed in the data dumps with RDF triples (e.g. by saying ``\textit{Place of Birth - is a - property}''). For each of the 89 domains, the following statistics on each domain's ontology were obtained.

\begin{itemize}[noitemsep]
\item The number of types and properties in the domain
\item The number of descriptions for each type and property
\item The number of property details for each type and property
\end{itemize}

\begin{figure}[h]
	\centering
	\caption{Scatterplots of \textit{complexity score} and \textit{triples count} for all slices; and without \textit{music}.}
	\includegraphics[width=\columnwidth]{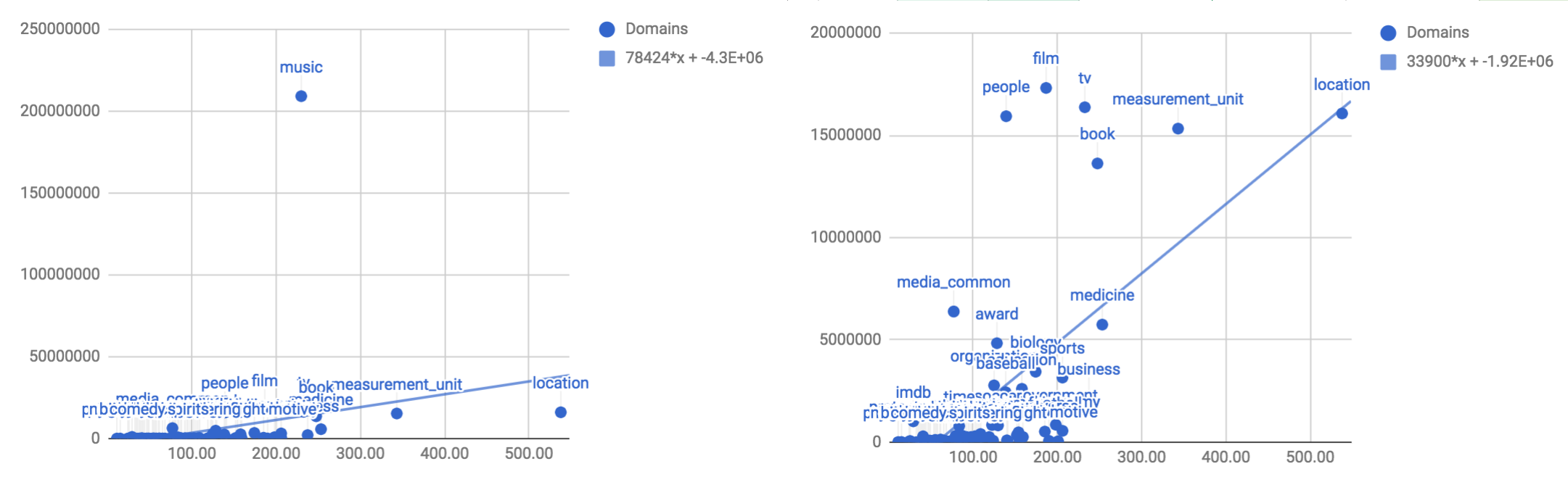}
	\label{fig6}
\end{figure}

A simple complexity score was calculated by obtaining the average number of descriptions and property details for each type and property in a domain. The Pearson's r correlation coefficient between the count of RDF triples and this complexity score for all domains was positively correlated at 0.2824, with the slope of a simple linear regression as 78,424.08 (see Figure \ref{fig6}). When the outlier Music slice is excluded, the correlation and slope become 0.6680 and 33,899.53 respectively. Although further work is needed to explore this research question, this small correlational study offers some promising initial results for further experimentation.

\section{Discussion}

This paper has explored select parts of the Freebase ontology and architecture to discover how they can be related to Bowker and Star's general findings on classification schemes. Three parts of the overall architecture were considered: the Freebase type system and its lack of inheritance and reliance on incompatibility, the implementation that allows representation of uncertainty in values, and the implementation of merging and splitting objects. In addition, a small correlational study was conducted to test a hypothesis based on a hunch motivated by Bowker and Star. To a large extent, many of the characteristics found in classification systems were also found in the ontology and architecture of Freebase. There are ``hacks'' and workarounds in the system that implement features to make it possible for Google to minimize the potential damage from embarrassing KPs and promote Google's features as an attractive destination for the end user. These ``security'' features are implemented in an environment of linked data, where users can hop back and forth between entities and their relationships with other entities. 

The Google Knowledge Graph is a proprietary system that powers the structured answers to queries delivered in Search, Home, and Assistant. Very little of the operations and implementation details of this KG are known to the public, yet it is tapped into increasingly frequently as the scope of virtual assistants and home assistants rises. By exploring these aspects of Freebase, it may be possible to develop an understanding of the Google KG, of which Freebase was a significant part. Specifically, the underlying structures (the ontology and architecture) that support the overall pipeline of delivering answers were explored rather than the particular facts that are represented in Freebase/KG.

Google's mission statement is to ``to organize the world's information and make it universally accessible and useful'' \cite{googlemission}. In the field of information studies, the term \textit{information} takes on vastly different definitions and uses depending on who provides the definition. \textit{Information} in the sense of Claude Shannon, Michael Buckland, Peter Adriaans, or Fred Dretske take on very different flavours from one another \cite{shannon1949mathematical,buckland1991information,adriaans2010critical,dretske1983precis}. What does Google mean by \textit{information}? According to the exploration of Google's KG through Freebase that was conducted here, \textit{information} is most closely aligned with Buckland's conceptualization of \textit{information-as-thing}, information in the form of data and documents \cite{buckland1991information}. By organizing the world's information, Google is expressing the searchable world in billions of RDF triples that conform to the KG system's implemented ontology and architecture. Associated with this view of information are the issues with the ontology/architecture that were mentioned in this paper. Representations of entities are classified according to a system of types that come with implicit incompatibilities (one cannot be \textit{X} and also be a \textit{Y}). Only certain kinds of uncertainty and ambiguous values are captured. The system must be able to readily merge and split data as more and more of it is amassed from various sources. Ontological work likely concentrates on subject matters that are deemed valuable for search traffic.

There is also the sense that Google talks about information, but actually means to say knowledge as seen in the very name for their datastore: the \textit{Knowledge} Graph. Google may be trying to invoke Buckland's \textit{information-as-knowledge} \cite{buckland1991information}. As Buckland notes, \textit{information-as-knowledge} ``is intangible: one cannot touch it or measure it in any direct way. Knowledge, belief, and opinion are personal, subjective, and conceptual. Therefore, to communicate them, they have to be expressed, described, or represented in some physical way...'' \cite{buckland1991information}. It is likely that Google's mission, through its Search service, Home appliance, and virtual Assistant, is more broadly ``to organize the world's information[-as-thing] and make it universally accessible and useful [as information-as-knowledge]''.

\section{Conclusion}

This paper has applied concepts from Bowker and Star's findings on classification systems to a popular company's once public database. Due to the incredible diversity of domains and features that can be found in the Freebase data dumps, only a small selection of relevant aspects could be examined here. A small correlational study was also conducted to obtain preliminary findings. Further research should be conducted by exploring additional aspects of the Freebase ontology and architecture and by conducting a more thorough experimental analysis of Freebase. This paper ends in the same way it started, by invoking the special phrase to say: ``OK Google, let's do more research.''

\newpage{}

\bibliography{biblio}
\bibliographystyle{plain}

\newpage{}

\section{Appendix A}

All of the Freebase domains have been clustered into one of three groups: Freebase Implementation Domains, OWL Domains, and Subject Matter Domains. The Freebase Implementation Domains include the triples that express the ontology and other helper domains that implemented Freebase on a technical level. OWL Domains include the triples based on the standard Web Ontology Language (OWL). Subject Matter Domains cover various domain areas on different subject matters in the world, from football to zoos.

\begin{longtable}{ | l | l | p{2.1cm} | r | r | r | }
\caption{Freebase Domains} \\ \hline
No. & Name & Domain & Triples & Total \% &  Group \% \\ \hline
\multicolumn{6}{|l|}{\textit{Freebase Implementation Domains}} \\ \hline
1 & common & /common/* & 1,429,443,085 & 45.658\% & 58.507\% \\ \hline
2 & type & /type/* & 788,652,672 & 25.191\% & 32.280\% \\ \hline
3 & key & /key/* & 149,564,822 & 4.777\% & 6.122\% \\ \hline
4 & kg & /kg/* & 30,689,453 & 0.980\% & 1.256\% \\ \hline
5 & base & /base/* & 24,063,303 & 0.769\% & 0.985\% \\ \hline
6 & freebase & /freebase/* & 11,259,415 & 0.360\% & 0.461\% \\ \hline
7 & dataworld & /dataworld/* & 7,054,575 & 0.225\% & 0.289\% \\ \hline
8 & topic\_server & /topic\_server/* & 1,010,720 & 0.032\% & 0.041\% \\ \hline
9 & user & /user/* & 912,258 & 0.029\% & 0.037\% \\ \hline
10 & pipeline & /pipeline/* & 547,896 & 0.018\% & 0.022\% \\ \hline
11 & kp\_lw & /kp\_lw/* & 1,089 & 0.000\% & 0.000\% \\ \hline

\multicolumn{6}{|l|}{\textit{OWL Domains}} \\ \hline
1 & type & rdf-syntax-ns\#type & 266,321,867 & 8.507\% & 78.520\% \\ \hline
2 & label & rdf-schema\#label & 72,698,733 & 2.322\% & 21.434\% \\ \hline
3 & domain & rdf-schema\#domain & 71,338 & 0.002\% & 0.021\% \\ \hline
4 & range & rdf-schema\#range & 71,200 & 0.002\% & 0.021\% \\ \hline
5 & inverseOf & owl\#inverseOf & 12,108 & 0.000\% & 0.004\% \\ \hline

\multicolumn{6}{|l|}{\textit{Subject Matter Domains}} \\ \hline
1 & music & /music/* & 209,244,812 & 6.684\% & 60.062\% \\ \hline
2 & film & /film/* & 17,319,142 & 0.553\% & 4.971\% \\ \hline
3 & tv & /tv/* & 16,375,388 & 0.523\% & 4.700\% \\ \hline
4 & location & /location/* & 16,071,442 & 0.513\% & 4.613\% \\ \hline
5 & people & /people/* & 15,936,253 & 0.509\% & 4.574\% \\ \hline
6 & measurement\_unit & /measurement \_unit/* & 15,331,454 & 0.490\% & 4.401\% \\ \hline
7 & book & /book/* & 13,627,947 & 0.435\% & 3.912\% \\ \hline
8 & media\_common & /media \_common/* & 6,388,780 & 0.204\% & 1.834\% \\ \hline
9 & medicine & /medicine/* & 5,748,466 & 0.184\% & 1.650\% \\ \hline
10 & award & /award/* & 4,838,870 & 0.155\% & 1.389\% \\ \hline
11 & biology & /biology/* & 3,444,611 & 0.110\% & 0.989\% \\ \hline
12 & sports & /sports/* & 3,158,835 & 0.101\% & 0.907\% \\ \hline
13 & organization & /organization/* & 2,778,122 & 0.089\% & 0.797\% \\ \hline
14 & education & /education/* & 2,609,837 & 0.083\% & 0.749\% \\ \hline
15 & baseball & /baseball/* & 2,444,241 & 0.078\% & 0.702\% \\ \hline
16 & business & /business/* & 2,134,788 & 0.068\% & 0.613\% \\ \hline
17 & imdb & /imdb/* & 1,020,270 & 0.033\% & 0.293\% \\ \hline
18 & government & /government/* & 852,785 & 0.027\% & 0.245\% \\ \hline
19 & cvg & /cvg/* & 841,398 & 0.027\% & 0.242\% \\ \hline
20 & soccer & /soccer/* & 820,410 & 0.026\% & 0.235\% \\ \hline
21 & time & /time/* & 791,442 & 0.025\% & 0.227\% \\ \hline
22 & astronomy & /astronomy/* & 556,381 & 0.018\% & 0.160\% \\ \hline
23 & basketball & /basketball/* & 519,652 & 0.017\% & 0.149\% \\ \hline
24 & american\_football & /american \_football/* & 483,372 & 0.015\% & 0.139\% \\ \hline
25 & olympics & /olympics/* & 400,927 & 0.013\% & 0.115\% \\ \hline
26 & fictional\_universe & /fictional \_universe/* & 349,147 & 0.011\% & 0.100\% \\ \hline
27 & theater & /theater/* & 320,721 & 0.010\% & 0.092\% \\ \hline
28 & visual\_art & /visual\_art/* & 310,238 & 0.010\% & 0.089\% \\ \hline
29 & military & /military/* & 292,533 & 0.009\% & 0.084\% \\ \hline
30 & protected\_sites & /protected \_sites/* & 288,788 & 0.009\% & 0.083\% \\ \hline
31 & geography & /geography/* & 256,768 & 0.008\% & 0.074\% \\ \hline
32 & broadcast & /broadcast/* & 256,312 & 0.008\% & 0.074\% \\ \hline
33 & architecture & /architecture/* & 253,718 & 0.008\% & 0.073\% \\ \hline
34 & food & /food/* & 253,415 & 0.008\% & 0.073\% \\ \hline
35 & aviation & /aviation/* & 187,187 & 0.006\% & 0.054\% \\ \hline
36 & finance & /finance/* & 131,762 & 0.004\% & 0.038\% \\ \hline
37 & transportation & /transportation/* & 112,099 & 0.004\% & 0.032\% \\ \hline
38 & boats & /boats/* & 108,763 & 0.003\% & 0.031\% \\ \hline
39 & computer & /computer/* & 106,986 & 0.003\% & 0.031\% \\ \hline
40 & royalty & /royalty/* & 92,787 & 0.003\% & 0.027\% \\ \hline
41 & library & /library/* & 86,249 & 0.003\% & 0.025\% \\ \hline
42 & internet & /internet/* & 80,426 & 0.003\% & 0.023\% \\ \hline
43 & wine & /wine/* & 79,520 & 0.003\% & 0.023\% \\ \hline
44 & projects & /projects/* & 79,102 & 0.003\% & 0.023\% \\ \hline
45 & chemistry & /chemistry/* & 72,698 & 0.002\% & 0.021\% \\ \hline
46 & cricket & /cricket/* & 67,422 & 0.002\% & 0.019\% \\ \hline
47 & travel & /travel/* & 56,297 & 0.002\% & 0.016\% \\ \hline
48 & symbols & /symbols/* & 56,139 & 0.002\% & 0.016\% \\ \hline
49 & religion & /religion/* & 54,887 & 0.002\% & 0.016\% \\ \hline
50 & influence & /influence/* & 53,976 & 0.002\% & 0.015\% \\ \hline
51 & language & /language/* & 53,588 & 0.002\% & 0.015\% \\ \hline
52 & community & /community/* & 50,164 & 0.002\% & 0.014\% \\ \hline
53 & metropolitan\_transit & /metropolitan \_transit/* & 47,777 & 0.002\% & 0.014\% \\ \hline
54 & automotive & /automotive/* & 46,543 & 0.001\% & 0.013\% \\ \hline
55 & digicams & /digicams/* & 42,188 & 0.001\% & 0.012\% \\ \hline
56 & law & /law/* & 37,606 & 0.001\% & 0.011\% \\ \hline
57 & exhibitions & /exhibitions/* & 37,434 & 0.001\% & 0.011\% \\ \hline
58 & tennis & /tennis/* & 34,853 & 0.001\% & 0.010\% \\ \hline
59 & venture\_capital & /venture \_capital/* & 27,410 & 0.001\% & 0.008\% \\ \hline
60 & opera & /opera/* & 26,630 & 0.001\% & 0.008\% \\ \hline
61 & comic\_books & /comic\_books/* & 25,529 & 0.001\% & 0.007\% \\ \hline
62 & amusement\_parks & /amusement \_parks/* & 22,880 & 0.001\% & 0.007\% \\ \hline
63 & dining & /dining/* & 21,297 & 0.001\% & 0.006\% \\ \hline
64 & ice\_hockey & /ice\_hockey/* & 17,275 & 0.001\% & 0.005\% \\ \hline
65 & event & /event/* & 14,783 & 0.000\% & 0.004\% \\ \hline
66 & spaceflight & /spaceflight/* & 14,238 & 0.000\% & 0.004\% \\ \hline
67 & zoo & /zoo/* & 13,226 & 0.000\% & 0.004\% \\ \hline
68 & meteorology & /meteorology/* & 12,432 & 0.000\% & 0.004\% \\ \hline
69 & martial\_arts & /martial\_arts/* & 12,065 & 0.000\% & 0.003\% \\ \hline
70 & periodicals & /periodicals/* & 9,424 & 0.000\% & 0.003\% \\ \hline
71 & games & /games/* & 9,024 & 0.000\% & 0.003\% \\ \hline
72 & celebrities & /celebrities/* & 8,815 & 0.000\% & 0.003\% \\ \hline
73 & nytimes & /nytimes/* & 7,537 & 0.000\% & 0.002\% \\ \hline
74 & rail & /rail/* & 7,431 & 0.000\% & 0.002\% \\ \hline
75 & interests & /interests/* & 5,345 & 0.000\% & 0.002\% \\ \hline
76 & atom & /atom/* & 5,199 & 0.000\% & 0.001\% \\ \hline
77 & boxing & /boxing/* & 4,282 & 0.000\% & 0.001\% \\ \hline
78 & comic\_strips & /comic\_strips/* & 4,234 & 0.000\% & 0.001\% \\ \hline
79 & conferences & /conferences/* & 2,495 & 0.000\% & 0.001\% \\ \hline
80 & skiing & /skiing/* & 1,949 & 0.000\% & 0.001\% \\ \hline
81 & engineering & /engineering/* & 1,546 & 0.000\% & 0.000\% \\ \hline
82 & fashion & /fashion/* & 1,535 & 0.000\% & 0.000\% \\ \hline
83 & radio & /radio/* & 1,385 & 0.000\% & 0.000\% \\ \hline
84 & distilled\_spirits & /distilled \_spirits/* & 1,055 & 0.000\% & 0.000\% \\ \hline
85 & chess & /chess/* & 558 & 0.000\% & 0.000\% \\ \hline
86 & physics & /physics/* & 449 & 0.000\% & 0.000\% \\ \hline
87 & geology & /geology/* & 353 & 0.000\% & 0.000\% \\ \hline
88 & bicycles & /bicycles/* & 313 & 0.000\% & 0.000\% \\ \hline
89 & comedy & /comedy/* & 120 & 0.000\% & 0.000\% \\ \hline
\end{longtable}

\end{document}